A 28nm Bulk-CMOS 4-to-8GHz <2mW Cryogenic Pulse Modulator for Scalable Quantum Computing

Joseph C Bardin[1,2], Evan Jeffrey[2], Erik Lucero[2], Trent Huang[2], Ofer Naaman[2], Rami Barends[2], Ted White[2], Marissa Giustina[2], Daniel Sank[2], Pedram Roushan[2], Kunal Arya[2], Benjamin Chiaro[3], Julian Kelly[2], Jimmy Chen[2], Brian Burkett[2], Yu Chen[2], Andrew Dunsworth[3], Austin Fowler[2], Brooks Foxen[3], Craig Gidney[2], Rob Graff[2], Paul Klimov[2], Josh Mutus[2], Matthew McEwen[3], Anthony Megrant[2], Matthew Neeley[2], Charles Neill[2], Chris Quintana[2], Amit Vainsencher[2], Hartmut Neven[4], John Martinis[2,3]

[1]University of Massachusetts, Amherst, MA, [2]Google, Goleta, CA,
[3]University of California, Santa Barbara, CA, [4]Google, Los Angeles, CA

While quantum processors are typically cooled to <25mK to avoid thermal disturbances to their delicate quantum states, all qubits still suffer decoherence and gate errors. As such, quantum error correction is needed to fully harness the power of quantum computing (QC). Current projections indicate that >1,000 physical qubits will be required to encode one error-corrected qubit [1]. Implementing a system with 1,000 error-corrected qubits will likely require moving from the contemporary paradigm where control and readout of the quantum processor is carried out using racks of room temperature electronics to one in which integrated control/readout circuits are located within the cryogenic environment and connected to the quantum processor through superconducting interconnects [2]. This is a major challenge, as the cryo ICs must be high performance and very low power (eventually <1mW/qubit). In this paper, we report the design and system-level characterization of a prototype cryo-CMOS IC for performing XY gate operations on transmon (XMON) qubits.

A XMON qubit [3] is a non-linear resonator comprised of a capacitor in parallel with a Josephson junction (see Fig. 1). The non-linearity results in anharmonic energy levels, with $f_{01}$ and $f_{01}-f_{12}$ typically 4–8GHz and 250MHz. The junction loop (SQUID) makes the XMON frequency-tunable. The XY, Z, and readout ports are used for microwave drive, frequency tuning, and readout, respectively. Due to the anharmonicity, a band-limited drive signal centered at $f_{01}$ applied to the XY port excites only the 0→1 transition. In this case, the qubit is represented by its two lowest energy levels: $|\psi\rangle = \cos(\theta/2)|0\rangle + \exp\{j\phi\}\sin(\theta/2)|1\rangle$, which can be interpreted as a point on the surface of the Bloch sphere (Fig. 1). The XY drive signal produces a deterministic rotation (gate operation) about an axis in the XY plane of the Bloch sphere, where the axis and angle of rotation are determined by the carrier phase and integrated envelope amplitude of the drive signal, respectively. The finite coherence time of a XMON (~0.1ms) motivates fast gates, but there is a tradeoff: the increased spectral width of fast pulses can drive the $f_{12}$ transition, causing errors. Thus, shaped pulses, typically Gaussian or cosine, are used. Pulse lengths and envelope amplitudes referenced to the XY port of the qubit are typically 10–30ns and 10–100μV.

A standard XMON control/readout system [4], as shown in Fig. 2, uses a number of AWGs and ADCs that are fast (1Gsps) and high-resolution (14b and 8b, respectively). We propose replacing the XY AWG with an IC of the architecture shown in Fig. 2, situated on the 3K stage of the system. The IC was designed to support carriers from 4–8GHz and consists of a pair of current-mode DACs that are controlled via a waveform memory and upconverted by a vector modulator. It was designed to operate at a physical temperature of 3K, where PDK models are unavailable: in addition to power draw and envelope accuracy, tolerance to increases in mismatch and threshold voltages [5,6] was considered during design. Fortunately, non-monotonicity and/or non-linearity in the DACs is tolerable for this application, since calibration would be carried out in a practical QC system.

As shown in Fig. 3, the envelope currents are generated by a DAC architecture optimized to produce symmetric waveforms. Each DAC contains 11 8b sub-DACs. Once triggered, a state-machine sequentially enables each sub-DAC, producing a monotonically increasing staircase of current. Once all DACs have been enabled, they are disabled in the reverse order from which they were enabled, resulting in a 22 clock-cycle long pulse. The nominal weighting of each sub-DAC is found from the derivative of the desired envelope. On-chip memory permits storage of 16 waveforms (instructions). The 8b weightings for each sub-DAC and the value of the reference current for each envelope generator ($I_N$) is reconfigurable on a waveform-by-waveform basis. The reference current, $I_P$, has 6b of resolution. The baseband currents are filtered and then upconverted using pair of passive mixers, whose differential outputs are transformer-coupled and combined as a single-ended signal. Each baseband DAC is connected to the corresponding mixer through a polarity switch, enabling full phase coverage. The transformer-coupled LO signals drive amplifier chains that were optimized as a tradeoff between power draw and frequency range. Tuning capacitors are incorporated on the bondpad side of each transformer to permit optimum coupling from 4–8GHz. For simplicity, no provision was made to null LO leakage, as it can be cancelled off-chip and was not viewed as a fundamental limitation. This functionality could be added in a future design.

A die micrograph appears in Fig. 7, which also shows the wirebonded IC and the module used for testing. The module was mounted on the 3K stage of a dilution refrigerator for testing with a qubit. The setup (Fig. 4) included provisions to inject a signal from a standard XY AWG, null-out LO leakage, and monitor the IC output. The IC was characterized at 300K using an oscilloscope connected to TP1 and functioned with $f_{LO}$ and $f_{CLK}$ exceeding the range of 4–8GHz and 0.5–3GHz, respectively. The power required to drive the inputs of the clock and LO hybrids was less than –20dBm at 2GHz and –10dBm at 5.6GHz, respectively. An example waveform for $f_{LO}$=5.6GHz and $f_{CLK}$=2GHz appears in Fig. 4. Here, the chip was initialized with 16 different waveforms, which were stepped through using the select lines.

Once cold, the qubit was tuned to 5.6GHz and settings for AT1 and PH1 were found to minimize idle $|1\rangle$ state occupation (i.e., due to LO leakage). Rabi experiments using the cryo-CMOS IC were next carried out. As described in Fig. 5, state probabilities were measured as a function of pulse amplitude when the qubit had been initialized to $|0\rangle$ and then driven by either one or two raised cosine pulses of varying amplitude. The amplitude was varied by sweeping the DAC reference current ($I_N$) for eleven different values of $I_P$. At each point, state probabilities were computed from 5,000 measurements. At 3K, the DACs producing $I_N$ were non-linear and non-monotonic, so the Rabi data are plotted versus pulse amplitude, measured at the XY monitor port using a spectrum analyzer. The results in Fig. 5 show the expected behavior: the maxima of the $|0\rangle$ and $|1\rangle$ state probabilities are consistent with separately measured $|0\rangle$ and $|1\rangle$ state readout error rates of 2.4% and 6.8%.

The fast-switching and phase-control features of the IC were evaluated using a three-gate experiment, consisting of (1) initializing the qubit to the $|0\rangle$ state, (2) applying an X-pulse to produce a rotation of $\theta_A$ about the X-axis, (3) applying a $\pi$-pulse with carrier phase $\phi_B$ to produce a rotation of $\pi$ radians about a vector at an angle of $\phi_B$ from the x-axis in the XY plane, (4) applying a second X-pulse to produce a rotation of $\theta_A$ about the X-axis, and (5) measuring the qubit state (see Fig. 6). This sequence was run for $\phi_B$ in (0, 2$\pi$) and $A_A$ such that $\theta_A$ was estimated to be in the range of 0 to $\pi$. Raised cosine envelopes were used for all pulses. The results, which show the IC can be used to perform coherent quantum control, appear in Fig. 6 alongside baseline measurements taken using standard qubit control electronics. The RMS error for the IC is <12%, which is excellent, given that no calibration was performed.

The cryo-CMOS IC is compared to a standard system in Fig. 6. Of particular importance is the low power, which was estimated by measuring the total AC and DC power required to continuously drive π-pulses. While the demonstration of a cryogenic quantum control interface dissipating <2mW is an important step towards the development of a scalable quantum control and measurement system, much research is still required to implement such a system. Future work could focus on calibration algorithms or the development of ICs for control of multiple qubits, qubit readout, or other related applications.

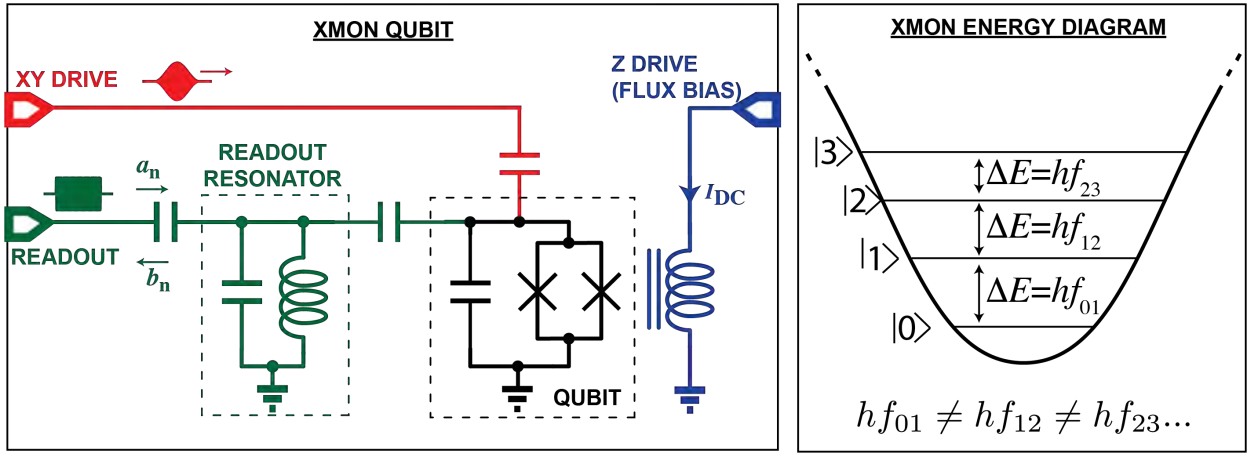

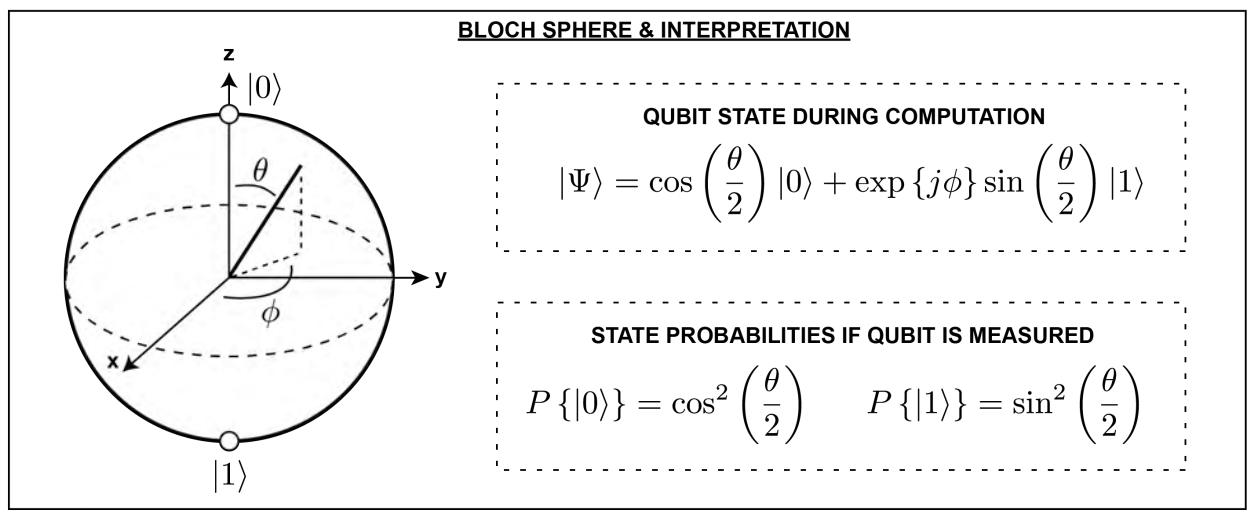

Figure 1: Transmon qubit schematic, energy diagram, and Bloch sphere representation. "X" schematic symbols represent Josephson Junctions.

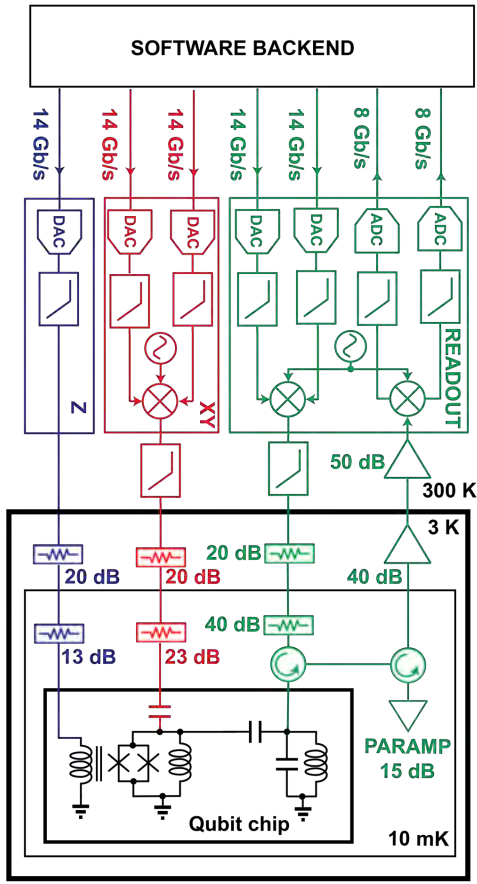 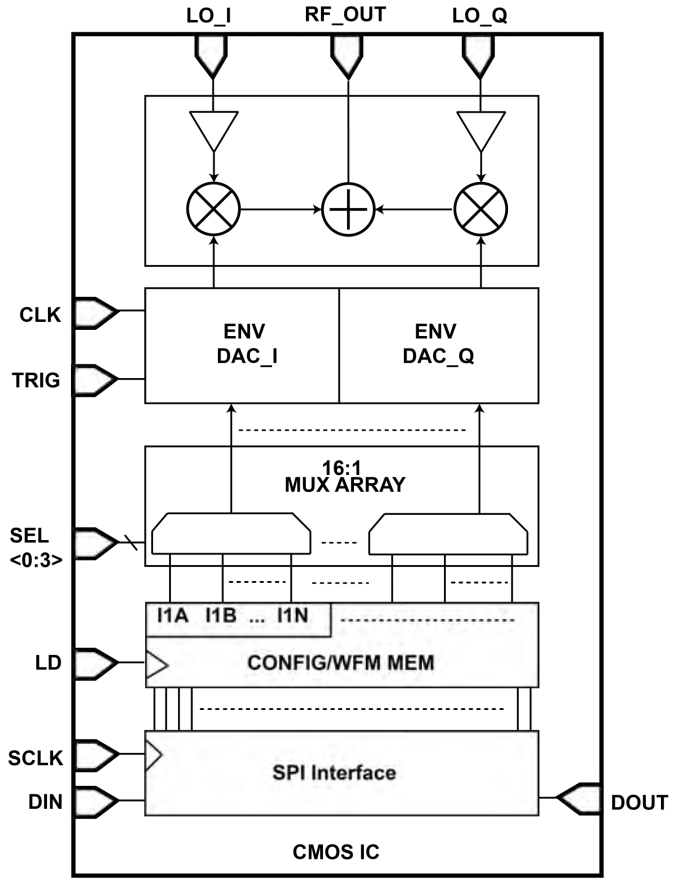

Figure 2: Standard qubit control/readout hardware and proposed XY Control IC

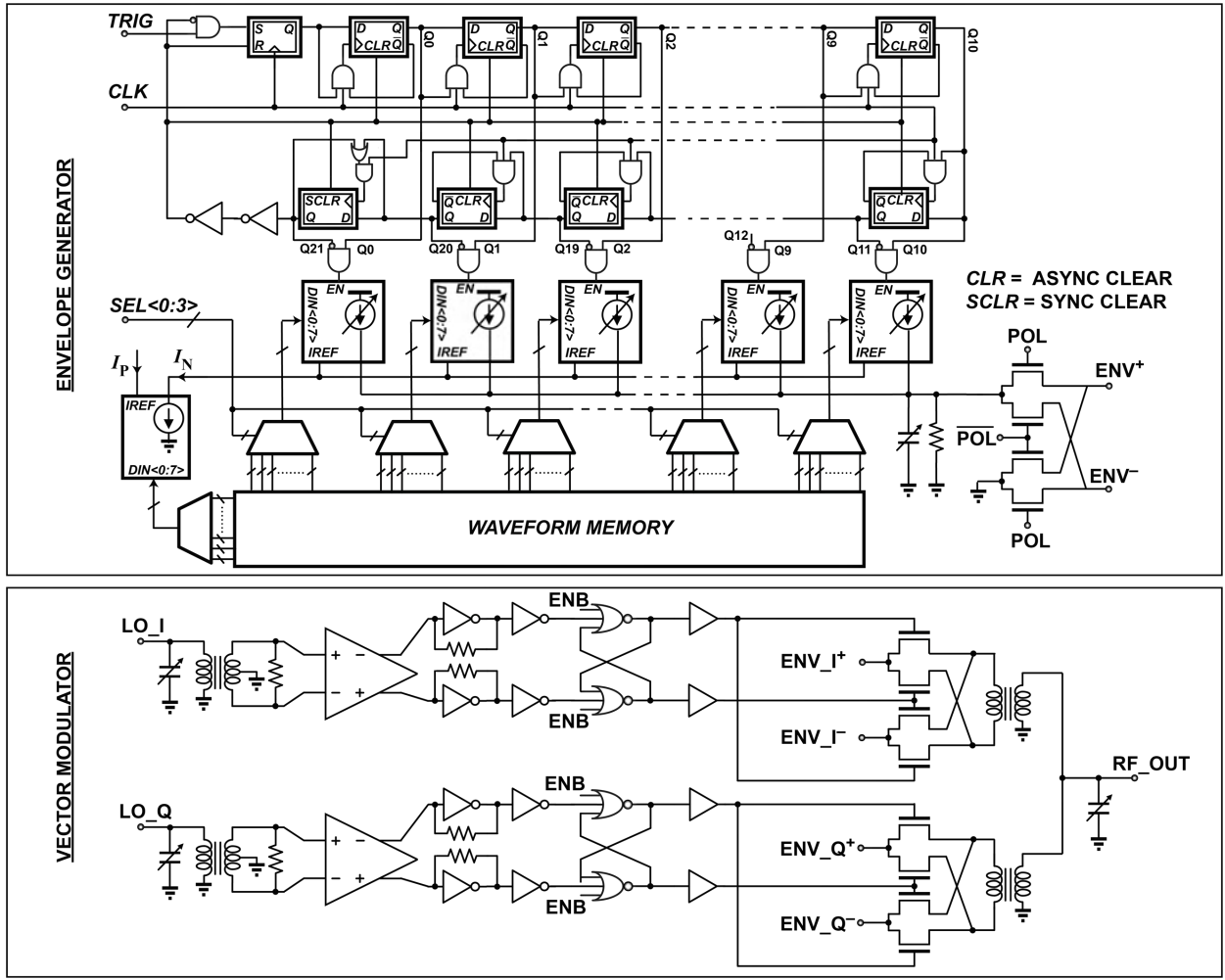

Figure 3: Schematic diagrams of current-mode envelope generator and vector modulator.

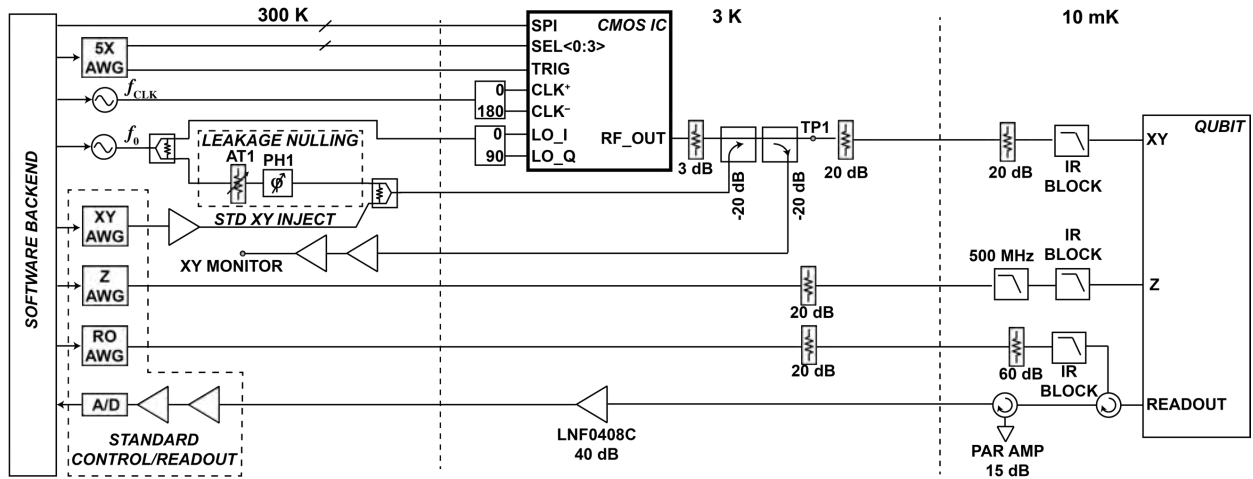

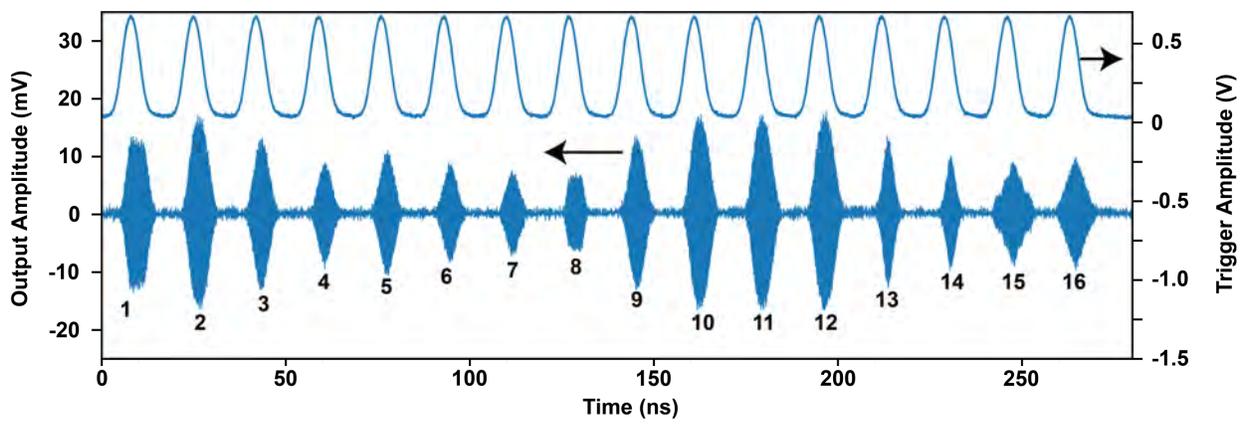

*Waveforms: 1: CRC #1, 2: RC #1, 3: RC #2, 4: RC #3, 5: RC #4, 6: RC#5, 7: RC #6, 8: SC #1, 9: RC #7, 10: RC #1, 11: RC#1, 12: RC #1, 13: GN #1, 14: GN#2, 15: SC #2, 16: TR#1*
*Abbreviations: (C)RC = (clipped) raised cos, SC = staircase, GN = Gaussian, TR = Triangular*

Figure 4: Experimental configuration and time domain waveforms, measured at room temperature.

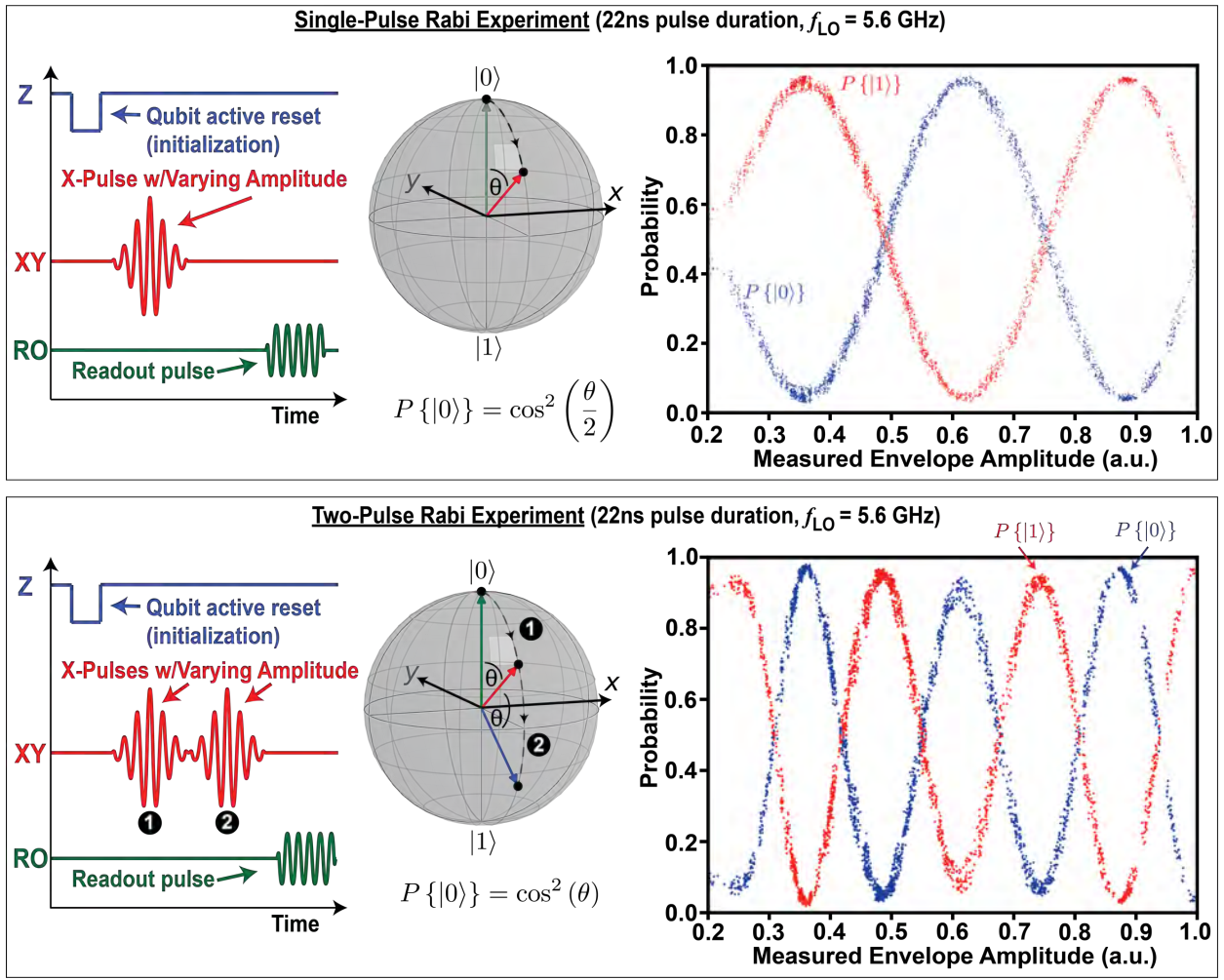

Figure 5: One- and two-pulse Rabi experiments. The envelope amplitude could not be measured below ~0.25 due to SNR.

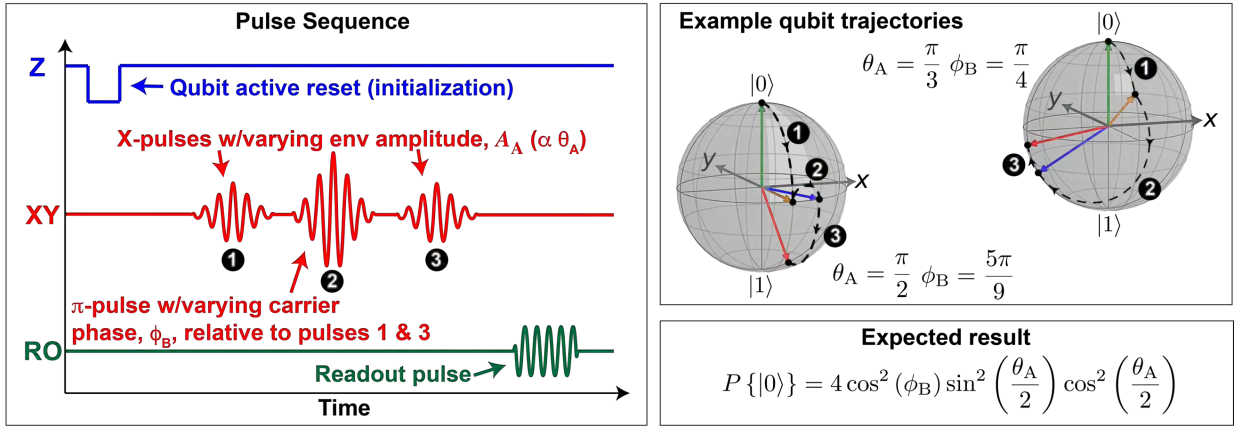

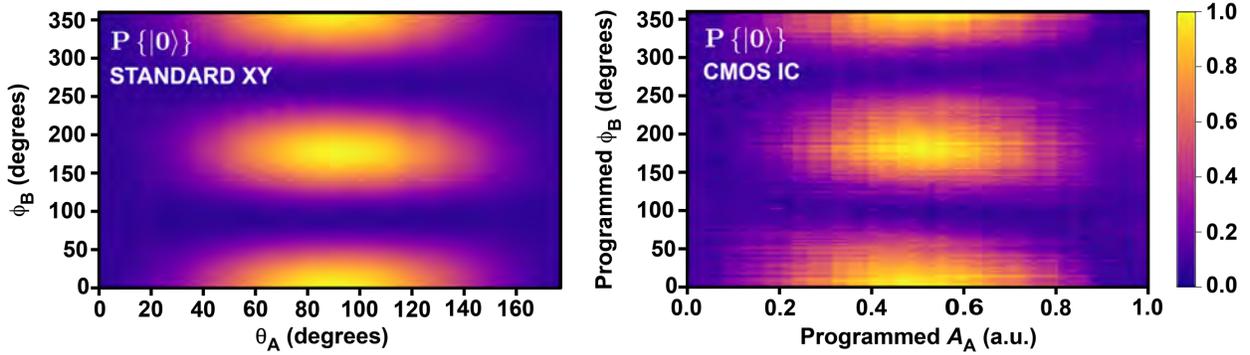

|  | Form Factor | Phys. Temp. | Update Freq. | Dig. Data Rate | π-Rabi P{|1⟩}* | 3-gate RMS err† | AC+DC Power |
|---|---|---|---|---|---|---|---|
| Standard XY | Rack Mount | 300 K | 1 Gsps | 28 Gbps | ~95% | 2.5% | > 1 W |
| This work | Integrated Circuit | 3 K | 1 Gsps | < 0.5 Gbps | ~95% | 11.7% | < 2 mW |

*In both cases, the measured probabilities were limited by the readout error rate
†Error defined as the difference between the ideal result and the measured result. The ideal result has been corrected to account for readout fidelity. X and Y axes for CMOS IC are uncalibrated.

Figure 6: Pulse sequence experiment and comparison to standard XY control architecture.

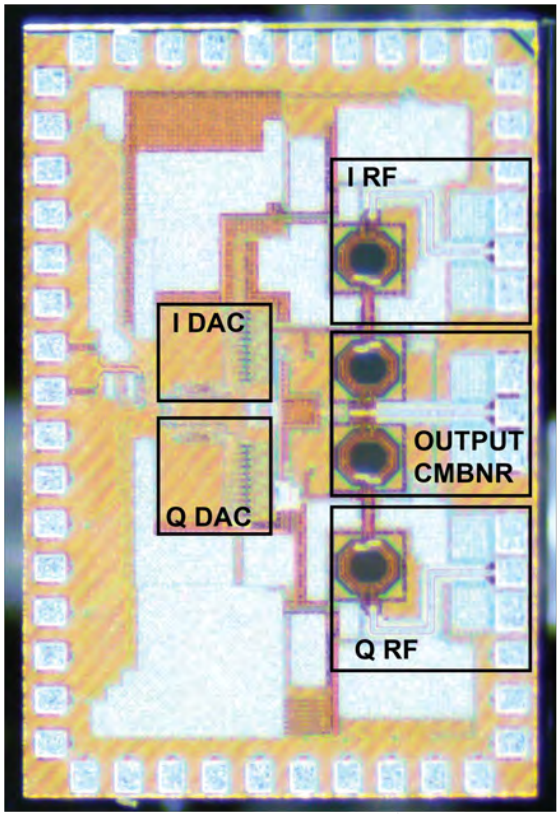
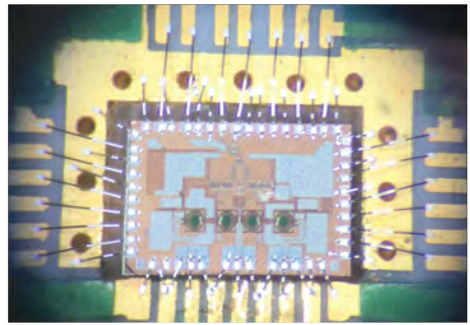
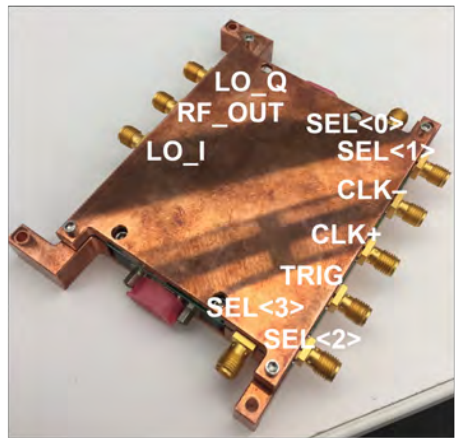

Figure 7: Die, PCB interface, and module photos. The integrated circuit measures 1mm by 1.6mm.